%% file: main.tex
\documentclass[10pt,conference]{IEEEtran}
\IEEEoverridecommandlockouts
\usepackage{cite}
\usepackage{amsmath,amssymb,amsfonts}
\usepackage[hyphens]{url}
\usepackage{graphicx}
\usepackage{textcomp}
\usepackage[T1]{fontenc}
\usepackage{xcolor}
\def\BibTeX{{\rm B\kern-.05em{\sc i\kern-.025em b}\kern-.08em
    T\kern-.1667em\lower.7ex\hbox{E}\kern-.125emX}}

\newcommand{\RN}[1]{%
  \textup{\uppercase\expandafter{\romannumeral#1}}%
}

\usepackage{graphicx}
\usepackage[flushleft]{threeparttable}
\usepackage{wrapfig}
\usepackage{amsmath,amssymb,amsfonts}
\usepackage{graphicx}
\usepackage{enumitem}
\usepackage{caption, booktabs}
\usepackage{booktabs}
\usepackage{lscape}
\usepackage{placeins}
\usepackage{textcomp}
\usepackage{multirow}
\usepackage{array}
\usepackage{listings}
\usepackage{color}
\usepackage{changepage}
\usepackage{tikz}
\usepackage{algorithm,algpseudocode}

\begin{document}

\title{TaskAllocator: A Recommendation Approach for Role-based Tasks Allocation in Agile Software Development}

% \IEEEauthorblockA{\textit{dept. of organization (of Aff.)} \\
% \and
% \IEEEauthorblockA{\textit{dept. of organization (of Aff.)} \\
% \and
% \IEEEauthorblockA{\textit{dept. of organization (of Aff.)} \\
% \and
% \IEEEauthorblockA{\textit{dept. of organization (of Aff.)} \\

\author{
\IEEEauthorblockN{1\textsuperscript{st} Saad Shafiq}
\IEEEauthorblockA{\textit{Johannes Kepler University} \\
%\textit{Johannes Kepler University}\\
Linz, Austria \\
saad.shafiq@jku.at% or 0000-0002-5901-1420
}
\and
\IEEEauthorblockN{2\textsuperscript{nd} Atif Mashkoor}
\IEEEauthorblockA{\textit{Johannes Kepler University} \\
Linz, Austria \\
atif.mashkoor@jku.at% or 0000-0003-3128-5427
}
\and
\IEEEauthorblockN{3\textsuperscript{rd} Christoph Mayr-Dorn}
\IEEEauthorblockA{\textit{Johannes Kepler University} \\
%\textit{Johannes Kepler University}\\
Linz, Austria \\
christoph.mayr-dorn@jku.at% or 0000-0001-9791-6442
}
\and
\IEEEauthorblockN{4\textsuperscript{th} Alexander Egyed}
\IEEEauthorblockA{\textit{Johannes Kepler University} \\
%\textit{Johannes Kepler University}\\
Linz, Austria \\
alexander.egyed@jku.at% or 0000-0003-3128-5427
}
}

\maketitle

\begin{abstract}
In this paper, we propose a recommendation approach -- {\it TaskAllocator} -- in order to predict the assignment of incoming tasks to potential befitting roles. The proposed approach, identifying team roles rather than individual persons, allows project managers to perform better tasks allocation in case the individual developers are over-utilized or moved on to different roles/projects. We evaluated our approach on ten agile case study projects obtained from the {\it Taiga.io} repository. In order to determine the TaskAllocator's performance, we have conducted a benchmark study by comparing it with contemporary machine learning models. The applicability of the TaskAllocator was assessed through a plugin that can be integrated with JIRA and provides recommendations about suitable roles whenever a new task is added to the project. Lastly, the source code of the plugin and the dataset employed have been made public.

\end{abstract}


\section{Introduction}
\label{sec:intro}

\input{intro.tex}

\section{Background}
\label{sec:background}

\subsection{Related work}
\label{subsec:related}

\input{related.tex}

\subsection{Why roles are important for task allocation?}
\label{subsec:motivation}
\input{motivation.tex}

\subsection{Structure of a task in project management platforms}
\label{subsec:Structure}
\input{Structure.tex}

\subsection{Use case scenario}
\label{subsec:example}
\input{example.tex}

\section{TaskAllocator}
\label{sec:approach}
\input{approach.tex}

\section{Case study}
\label{sec:case}
\input{casestudy.tex}

\section{Research Artifacts}
\label{sec:Usability}
\input{usability}

\section{Threats to validity}
\label{sec:ttv}
\input{ThreatsToValidity.tex}

\section{Conclusion}
\label{sec:conclusion}

\input{conclusion.tex}

\section*{Acknowledgment}
The research reported in this article has been partly funded by the LIT Artificial Intelligence Lab and the LIT Secure \& Correct Systems Lab supported by the state of Upper Austria.

\bibliographystyle{IEEEtran}
\bibliography{TaskAssignment-Literature}
\end{document}

%% file: intro.tex
Distributed agile software development (DASD) aims to combine the benefits of diversification that could possibly improve the knowledge gain from geospatial experiences of team members with rapid and continuous delivery. Despite the benefits one could yield from DASD, it comes with inevitable challenges, which could affect the overall productivity of the project~\cite{Alzoubi2015d}. Some of those challenges include tasks (e.g., user stories and issues) allocation, poor communication among stakeholders, language barriers, and incompatible time-zones~\cite{Shrivastava2015}. The lack of knowledge of project managers about team members due to their scattered locations, varying and diverse time zones, and social and cultural conflicts also lead to poor tasks allocations, thus impacting project velocity~\cite{Aslam2018}. Imtiaz et al.~\cite{Imtiaz2017} state that an effective tasks allocation strategy significantly leads to optimal decisions, thus benefiting the project. Tasks are allocated by project managers to respective team members based on their roles, workload, and expertise. Coupled with the aforementioned risks, an optimal allocation could easily take a considerable amount of time. If not performed carefully, poor allocation could possibly result in greater task times, decrease in the quality of software, and cost and budget overruns.

Tasks allocation in industry is usually done manually by project managers. The highly subjective nature of this activity may often lead to inefficient, error-prone, and poor quality decision making~\cite{Barcus2008}. Barcus et al.~\cite{Barcus2008} conclude in their study that unsupported decision making especially when it comes to distributed teams is far from desirable due to the lack of information sharing and performance uncertainties among project managers. %For instance, a project manager in the US office will not be able to manage a team in Asia office without having briefed regarding the teams. 
What is desirable here is an automated approach for tasks allocation reducing the overall manual effort of triaging tasks and allowing project managers to take better decisions.

The rapid advancement in machine learning (ML) techniques has led to promising results in various fields. For example, deep learning and reinforcement learning -- a subset of ML -- are now being utilized in software for automotive industry~\cite{Falcini2017}, image recognition~\cite{Chen2018b}, software cost estimation~\cite{Choetkiertikul2019}, and industrial production assembly lines~\cite{Shafiq2020a}. These ML techniques complement the effectiveness of the processes and lead to significant improvements in cutting down project costs. In software engineering, ML techniques are particularly helpful in the bug triaging problem~\cite{Anvik2006,Mani2018,Systems2018,Choquette-choo2019,Yadav2019}, i.e., identifying potential developers who can fix the bug.

In this paper, we propose to leverage the power of ML models in order to infer knowledge and information regarding tasks allocation from previous project history. Using this knowledge and information, our proposed approach (TaskAllocator) predicts the suitable team role for a newly added task to the project. We believe that, like bug triaging, allocating a new task to the appropriate team role can also be facilitated using prediction models. Predicting the role over an individual is particularly useful in case of a high turnover, which is common in the agile project development~\cite{DeMelo2013}. Additionally, this helps the uniform distribution of knowledge within the team. Consequently, the proposed approach helps in achieving improved performance and quality of the project in the long run by retraining the TaskAllocator on the data as the project matures.

The main contribution of this paper is a recommendation approach for \emph{role-based} tasks allocation to assist project managers. The TaskAllocator is evaluated on individual projects as well as in a cross-project setting. The results show that the TaskAllocator is able to successfully predict the role for an incoming or modified task with an accuracy of 69.3\%, inferring 2 out of 3 recommendations as correct. Moreover, we perform a comprehensive benchmark study that shows how the TaskAllocator model fares as compared to other contemporary ML and neural network (NN) models. Lastly, the contributions of this study also include a prototype implementation of the TaskAllocator along with a publicly available dataset. Although we employed agile projects as subjects in this paper, we believe that our approach is equally applicable to non-agile projects as well.

%Existing studies focus on triaging at the individual level, while TaskAllocator is specifically designed to predict coarse-grained roles which makes this study as first and less likely to be compared. Please note that due to this fundamental difference, a direct baseline comparison is unfeasible, however, this study could be a starting point for role based recommendation approaches.

The rest of the paper is organized as follows: Section~\ref{sec:background} describes the background for this study including related work and the need for roles for tasks allocation. %Section~\ref{sec:Structure} discusses the structure of a task. The motivation to conduct this study is described in Section~\ref{sec:motivation}. 
Section~\ref{sec:approach} demonstrates the TaskAllocator. Section~\ref{sec:case} further elaborates the presented approach using a case study. Section~\ref{BS} describes the benchmark study, which we conducted in order to determine the performance of the employed model. The research artifacts and practical implications of the proposed approach are discussed in Section~\ref{sec:Usability}. Section~\ref{sec:ttv} discusses the threats to validity of this study. The paper is concluded in Section~\ref{sec:conclusion}.

%% file: related.tex
Task allocation is an active area of research in agile software development. For example, Aslam et al.~\cite{Aslam2018} have proposed theoretical frameworks to achieve efficient task allocation based on technical preferences, expertise, and current workload of team members. 

Similar to task allocation, Yadav et al.~\cite{Yadav2019} proposed a two-staged developer expertise score (DES) for bug triaging. In this work, researchers use multiple criteria-based metrics including developers expertise, average bug fixing time, versatility score, and the priority of the bug.

A semi-automatic approach has been proposed by Anvik et al.~\cite{Anvik2006} dealing with the assignment of bug reports to developers. The approach comprises of a machine learning algorithm employed on open bug repositories to identify the developer having satisfactory expertise who can be assigned to fix the bug. 

Baysal et al.~\cite{Baysal2009} proposed a theoretical framework for automatic assignment of bugs to experts. The framework recommends the most suitable developers by inferring their expertise, preferences, and priority of the bug report.

Asri et al.~\cite{Asri2018a} proposed an approach to classify the core and peripheral developers in open-source software (OSS) projects. The approach used K-means clustering on social network analysis (SNA) metrics to determine the collaboration among the developers and distinguish among them based on core contributions.

All of these aforementioned studies focus on a fine-grained analysis of development teams (identifying individuals) and do not address the need of identifying the team roles in the first place. Our study, on the other hand, aims at a \emph{coarse-grained} analysis of teams where there is a substantial need to determine suitable roles not only in order to expedite the incoming task allocation process, but also to be compliant with project's and organization's goals such as uniform distribution of knowledge among team members. In theory, team roles can also be deduced from the identified individuals, however, good and self-managing agile teams possess multi-skilled individuals -- also known as generalists~\cite{Moe2013,belling2020agile} -- who keep on changing roles as required. Therefore, in practice, predicting team roles based on identified individuals is not a feasible solution. %The need for generalists is also corroborated by the studies~\cite{Moe2013,belling2020agile} indicating the immense need for cross-training within teams in order to improve the development process. Therefore, recommending the individuals directly is not appropriate in such cases.

%% file: motivation.tex
Task allocation requires careful planning and assessment because if the task is assigned to an individual who is not optimal for the task or leaves the project in the middle, the re-assignment may then take a considerable amount of time~\cite{Anvik2011}. Task allocation becomes even more complicated in geographically distributed teams where there is a lack of direct dyadic communication, hence poor distributed collaboration~\cite{Pichler2007}. Moreover, some developers possess expert knowledge on the particular module, code, or artifact -- also known as ``mavens''~\cite{Cetin2020}. Losing a team member, such as a maven, in the middle of a project would greatly impact the team productivity. In reality, this is often the case as turnover rate is quite high in agile projects~\cite{DeMelo2013}. Therefore, there is a need for organizations to provide cross-training within teams, thus producing generalists who can easily adapt to changing dynamics of incoming tasks. %These costs include separation expenses such as conducting exit interviews, administration process, recruitment and orientation of new member, and the time the new member takes before becoming completely productive. As this turnover rate is high in agile projects~\cite{DeMelo2013}, 
Additionally, spreading knowledge of various modules among team members is essential for managers in order to avoid such situations. Hence, unlike existing approaches, instead of identifying a particular individual, identification of the role, which can quickly become operational for the task would be more beneficial. This not only provides managers the flexibility to distribute knowledge within a team, but also enable to tailor the task allocation process in line with the project's and organization's goals. As also noted by other researchers, e.g.,~\cite{Ahmad2011,Lings2007}, determining proper roles and responsibilities and aligning them with corresponding tasks results into an effective and successful distributed collaboration.

%% file: Structure.tex
An agile project consists of a multitude of artifacts such as user stories (and their tasks thereof), and issues. The structure of these artifacts, however, may vary from one project management platform to another. For simplification, we will refer to user stories, tasks, and issues as ``tasks'' for the rest of the paper. A task ideally contains all the relevant information essential to complete it including textual as well as categorical data. The typical textual data in a task includes title, description, comments, date created, etc. Categorical data, on the other hand, includes priority, story point (in case of a user story), severity, component, reported by, assigned to, status, etc.

\begin{figure}[htbp]
  \centering
  \includegraphics[width=1.0\linewidth]{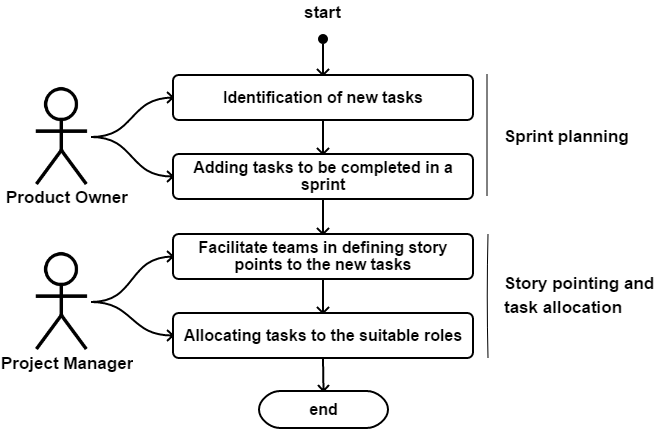}
  \caption{The tasks triaging process}
  \label{USlifecycle}
\end{figure}

If the task consists of a user story, it is first identified and checked for feasibility by the product owner. The task may then be decomposed into sub-tasks, if necessary. The task is then added to the product backlog by the product owner. At the start of each sprint, tasks that need to be completed in this sprint are identified. These tasks are then assigned story points, which determine their sizes. The project manager then decides the suitable individual/role to assign the task. On the other hand, if the task consists of an issue, it is either reported by the product owner or by the customer. The support staff then analyse the issue and affirms its replicability. Based on the type, severity and priority, the project manager assigns the task to the appropriate individual/role in the team. However, in the case of a geographically distributed team, this task assignment process is nontrivial. The  typical steps involved in the triaging process of tasks are illustrated in Fig.~\ref{USlifecycle}.

%% file: example.tex
Lets assume a company $X$ has a globally distributed software development team working on multiple projects. For the sake of simplicity, each project has one project manager and one product owner:

\begin{itemize}
    \item A new task $A$ is created by the product owner and must be included in the incoming sprint 
    \item The agile project manager has to decide which role is most suitable for the task $A$
    \item The most suitable role for $A$ is ``Front-end developer''
    \item The developer $John$ has the most expertise for $A$
\end{itemize}

Since John is a maven, the existing state of the art approaches, as discussed in Sec.~\ref{subsec:related}, will recommend John as the most suitable individual for the task. However, this may not be an ideal solution for multiple reasons, e.g., John may be busy with other important tasks or less interested in this task, the manager wants other team members to have more knowledge of the task, or even John is no more part of the team oblivious to the recommendation system. Hence, our approach recommends roles at a higher abstraction level allowing project managers make allocation decisions suitable for the situation, and compatible with project's and organization's goals. 

%Our approach takes into account the history of tasks of the project and its assigned roles, and recommends the role on a coarse-grained level that is most suitable for the new task $A$. Please note that our approach is only applicable for projects that have evolved over a significant time period and have considerable number of completed sprints.

%% file: approach.tex
\begin{figure}[htbp]
  \centering
  \includegraphics[width=1.0\linewidth]{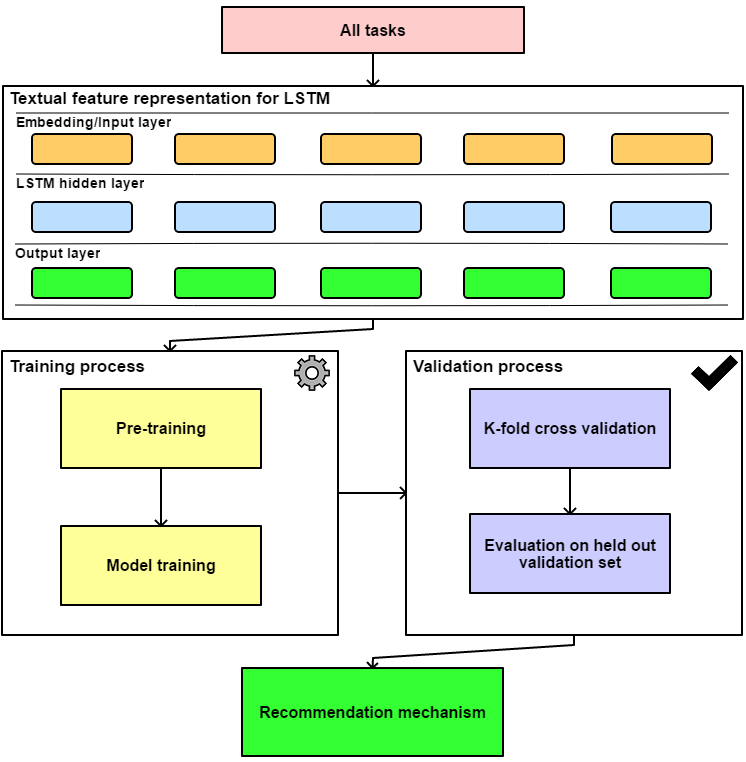}
  \caption{Architecture of the TaskAllocator}
  \label{Approach}
\end{figure}

  We propose an approach named TaskAllocator in order to assist project managers in allocating the newly added or the modified task to the most suitable role in the team. TaskAllocator learns from the textual features of previous tasks allocations and predicts the approximated role for the incoming tasks. One of the distinctive features of the TaskAllocator is its flexible architecture. Currently, the TaskAllocator is exploiting long short-term memory (LSTM)~\cite{Choetkiertikul2019} for predictions because (as shown later in the paper) it gives better results than the contemporary ML models. However, this component of the architecture is easily replaceable by any other ML model, which performs better than LSTM in the future or in another context. Fig.~\ref{Approach} shows the overall architecture of the TaskAllocator.

As an agile software development team consists of multiple but common roles, we decided to consider task allocation as a multi-class classification problem. As aforementioned, the TaskAllocator leverages the LSTM architecture, which is a variant of artificial recurrent neural network (RNN) and is often used for text classification problems. We use LSTM to learn the textual features (extracted from the title) in a task, train on these features, and predict the most suitable roles for the given text (title of the task). Please note that the TaskAllocator assumes that the tasks within a project are independent and only assigned to an individual instead of a group.

\subsection{Pre-training}\label{PT}
Pre-training is a process of initializing weights with the set of weights obtained from a previously trained neural network. This process helps the model in such a way that it does not have to learn and train from scratch (initialized with randomized weight values). We employed pre-trained vectors for the model to learn the domain ontologies (concepts and their relationships), and vocabulary before hand. The pre-trained vectors are provided by Efstathiou et al.~\cite{Efstathiou2018}. The vectors in the pre-trained dataset are taken from the well-known repository StackOverflow\footnote{{\url{https://stackoverflow.com/}}}, which covers the vocabulary of words commonly employed by developers in the software engineering domain.

\subsection{Model training}

The LSTM variant provided by Keras\footnote{\url{https://keras.io/}} has been employed for implementing the training process. Keras is an API -- written in the python language\footnote{\url{https://www.python.org/}}  -- comprised of various ML libraries including implementation of well-known deep learning techniques. The embedding tokens obtained from pre-training were then used as input to the LSTM model. The LSTM model learns the tokens against the corresponding label and outputs one class from the set of  classes. %Although, LSTM or any other deep learning model, in general, requires a lot of training data in order to be able to make accurate predictions, however, we aim to evaluate how the model performs on a considerably scarce dataset. 
While LSTM or the similar NN-based models tend to learn by optimizing loss, i.e.,  the cost function, for multi-class classification problems, categorical cross-entropy is used. This loss metric helps the model to distinguish between the two discrete probability distributions. The categorical cross-entropy loss function is computed using the following sum:
\begin{align}
\textbf{Loss} = -\sum_{c=1}^{N} y\textsubscript{s,c}\cdot log (\hat{y}\textsubscript{s}) \tag*{}\\\tag*{}
\end{align}

where $N$ is the number of classes in the model output, $y\textsubscript{s}$ is the target value of class $c$ for the $s$-th sample, and $\hat{y}\textsubscript{s}$ is the model output of class $c$ for the $s$-th sample. The minus sign refers to the decrease in loss as the values, i.e., target and model output values, get closer to each other.

\begin{figure}[htbp]
  \centering
  \includegraphics[width=0.75\linewidth]{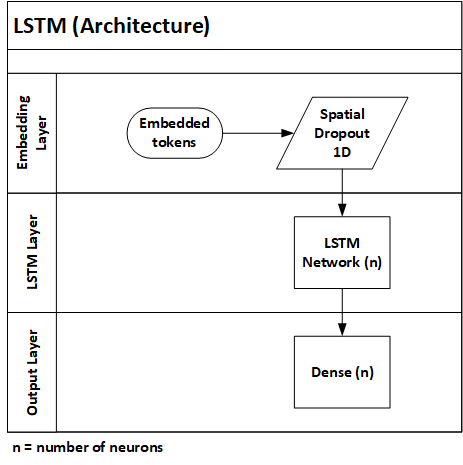}
  \caption{Architecture (LSTM)}
  \label{LSTM}
\end{figure}

Fig.~\ref{LSTM} shows the LSTM network architecture employed in this study. It comprises of three layers:
\begin{enumerate}
    \item The embedding layer contains the pre-trained vectors and the input vectors. It is followed by a spatial dropout 1D, which removes the one dimensional features from the input leading to learning of features with higher dimensions. 
    \item The LSTM layer contains the conventional LSTM architecture implemented in Keras. 
    \item The output layer contains the dense NN layer yielding classes (roles).
\end{enumerate}

\subsection{Validation process}

The validation process comprises of a validation set obtained by splitting the dataset into two parts: a train set used in the training of the LSTM model and a validation set used to evaluate its performance. The trained model is then evaluated using K-fold cross validation. K-fold cross validation is a method to evaluate the generalizability of the model by splitting the dataset into K folds. The model is then trained on K-1 folds and tested on the remaining fold. This process is repeated K times. After hyper-parameters\footnote{Hyper-parameters are set of variables that help in manipulating the behavior of a machine learning model.} optimization, the model, which performed better in K-fold cross validation, was selected. We further evaluated it on the held-out validation set as the model was not familiar to this set during the training process. The validation set contains randomized samples selected within the dataset. Accuracy -- the most commonly employed metric in order to evaluate the performance of classification problems~\cite{Silva-Palacios2017} -- is also employed in this paper. The accuracy metric represents the prediction capability of the model for all Classes N by computing true positives (TP), true negatives (TN), false positives (FP) (Type-\RN{1} error), and false negatives (FN) (Type-\RN{2} error).

\begin{align}
\textbf{Accuracy} = \sum_{x=1}^{N}\frac{TP\textsubscript{x} + TN\textsubscript{x}}{TP\textsubscript{x} + TN\textsubscript{x} + FP\textsubscript{x} + FN\textsubscript{x}} \tag*{}\\\tag*{}
\end{align}

% We further analyzed the performance of the model for each project in the validation set.

\subsection{Recommendation mechanism}
The TaskAllocator considers the previously assigned tasks in a project to assignee and generates a potentially suitable role for a new incoming task. The approach leverages requirements written in natural language, such as text in titles of tasks, to predict the befitting roles in an agile team. The text in titles are converted to sequence vectors, which later become multidimensional features for the training of an ML model as explained in the aforementioned training process. If a predicted role is not present in the project, our approach further recommends the role, which has highest confidence and presence in the project. We categorize team roles into a smaller, more generic group of roles as class $x$ based on the obtained team roles with similar job descriptions in the data.

\begin{algorithm}
\caption{TaskAllocator}
\label{Algo}

\begin{flushleft}
%\vspace{1em}
\textbf{Input} 

Task = \{task\textsubscript{1}....task\textsubscript{n}\}

Title per task \\

\textbf{Output} \[ Role^\prime(x) = [Role^\prime : task\implies task^\prime] \]

\textbf{Begin}

\begin{algorithmic}[1]

\State H = RandomSampling(Task,Role)

Random sampling of dataset based on Role

\State T\textsubscript{i} = Train ML model(H)

Train ML model with specified hyper-parameters

\State V\textsubscript{i} = Validate(T\textsubscript{i})
    \If{predicted role = true role}
   \State TruePredictions + 1    
\EndIf
\State TotalPredictions + 1

\State Compute Accuracy

\State M = Trained model with high V\textsubscript{i} accuracy

    \State  \[ Role^\prime(x) = \sum_{n=1}^{n} M\textsubscript{n}(task^\prime)\]
   
\If{$ Role^\prime(x) $ is not in the project}

    \State return $ Role^\prime(x) $ which has highest confidence and also exist in the project   
\Else{}
    \State return $ Role^\prime(x) $
\EndIf

%\EndWhile
\end{algorithmic}
\textbf{End}
\end{flushleft}
\end{algorithm}

% \vspace{2cm}
% \textcolor{red}{The Algorithm~\ref{NLPAlgo} of NLP4RP approach begins with computing rank from the importance and complexity factors as described in Section~\ref{SecRP}. Then, applying random sampling to the dataset followed by a split into K folds.} 
Following are the main steps of the TaskAllocator as presented in Algorithm~\ref{Algo}:

\begin{itemize}
	
	\item Line 1 - Generate random samples (H)
	
	\item Line 2 - Train (T) the ML model on the training set %T\textsubscript{i}
	
	\item Line 3 - Validate (V) the trained model on the validation set %\textsubscript{i}
	
	\item Line [4-7] - Compute the number of true predicted roles by comparing them with true roles in the validation set
	
	\item Line 8 - Compute the average accuracy
	
	\item Line [9-10] - Resultant Trained Model (M) demonstrating the high validation accuracy
	
	\item Line [11-12] - If the predicted role generated by the function $Role^\prime(x)$ is not in the project (only valid in cross-project setting) then return the role with highest probability in the project
	
	\item Line [13-15] - Else return the predicted role generated by the ML model corresponding to the newly added or modified task ($task^\prime$)
	
\end{itemize}

% The composition of hidden layers in the three models along with the pretrained model is provided in Table~\ref{Model Structure}.

% \begin{table}
% \centering
% \caption{Models Structure}
% \begin{tabular}{|c|c|c|}
% \hline
% Model & Hidden Layers& Pretrained Model \\
% \hline
% CNN& 3 & Yes\\
% \hline
% USE& 1 & Yes\\
% \hline
% LSTM& 1 & Yes\\
% \hline
% \end{tabular}
% \label{Model Structure}
% \end{table}

%% file: casestudy.tex
In order to extract data from software development projects with substantial representation of software team roles, we initially assessed multiple well-known repositories including JIRA\footnote{\url{https://www.atlassian.com/software/jira}}, Github\footnote{\url{https://github.com/}}, and Bitbucket\footnote{\url{https://bitbucket.org/}}. However, in these repositories the information regarding roles of the team members was accessible only to project administrators; hence not publicly available. Therefore, we have opted for another well-known {\it Taiga.io}\footnote{\url{https://taiga.io/}} repository as our primary source for data extraction. Various studies, such as~\cite{Shafiq2017,Shafiq2019}, have already employed this repository in the past for the purpose of identifying communication patterns within agile teams. {\it Taiga.io} is a project management platform for agile teams and has a total of 402,771\footnote{Checked in December 2020} members actively using the platform. {\it Taiga.io} also provides access to real public projects along with access to their tasks, user stories, issues, and team roles.

\subsection{Data extraction}\label{DE}
The data is extracted using the REST API provided by {\it Taiga.io}, which is comprised of title, description, and team roles for each task. The dataset contains 1,226 tasks (user stories, and issues combined). In order to select the most appropriate projects for our study, we have defined a project selection criteria. The criteria ensures a systematic selection of projects with sufficient maturity needed for the study. We searched for the most active and liked projects in the repository (search filters provided by the repository). The search for projects was stopped when projects appeared to show least activity and least number of likes (0) on the current page. Moreover, the project selection criteria originally stems from the work of Shafiq et al.~\cite{Shafiq2017} and has been adapted for this work after a series of discussions among the authors of this paper who are experienced in agile software development. Following is the project selection criteria:  

%\subsection{Project Selection Criteria} \label{PSC}
%The project selection criteria is defined to include specific projects complying to the research goals. Following are the criteria for the project selection:

\begin{enumerate}

\item The project must have at least 100 tasks, user stories, and issues combined.
\item The project must have at least 5 team members.
\item The project must have at least 5 sprints.
\item The project must have more than 3 team roles.

\end{enumerate}

The process of extraction yielded a total of 10 projects that met the aforementioned criteria and were subsequently used in this study.

\subsection{Data preprocessing}\label{DP}
The dataset obtained from {\it Taiga.io} is then preprocessed. First, we removed the HTML tokens from the text in titles. Then, we converted the text to lower-case and also replaced symbols with spaces. In order to further improve the training process, we have used the stop words corpus for English -- provided by natural language toolkit\footnote{\url{https://www.nltk.org/}}(NLTK) -- on the text for the removal of stop words. The text of the titles of tasks of all the projects was tokenized using the Keras tokenizer method\footnote{\url{https://keras.io/api/preprocessing/text/}}. A tokenizer converts each text into sequence of integers, and maintains the morphological relationship and context among words. It then stores the vocabulary index based on the frequency of words in the text. These tokens ultimately become features of the machine learning model. On the other hand, the team roles (labels) naturally represented as categorical data are converted into integers using one hot encoding method~\cite{Lantz2019}\footnote{One hot encoding refers to the process of converting categorical variables into binary vectors.}. Based on the role distribution in projects, we generalized the roles into broader categories. The total number of generalized roles is seven. Fig.~\ref{RD} shows the roles distribution of the held-out validation set. Following is the description regarding generalized roles:

\begin{figure*}
  \centering
  \includegraphics[width=.75\linewidth]{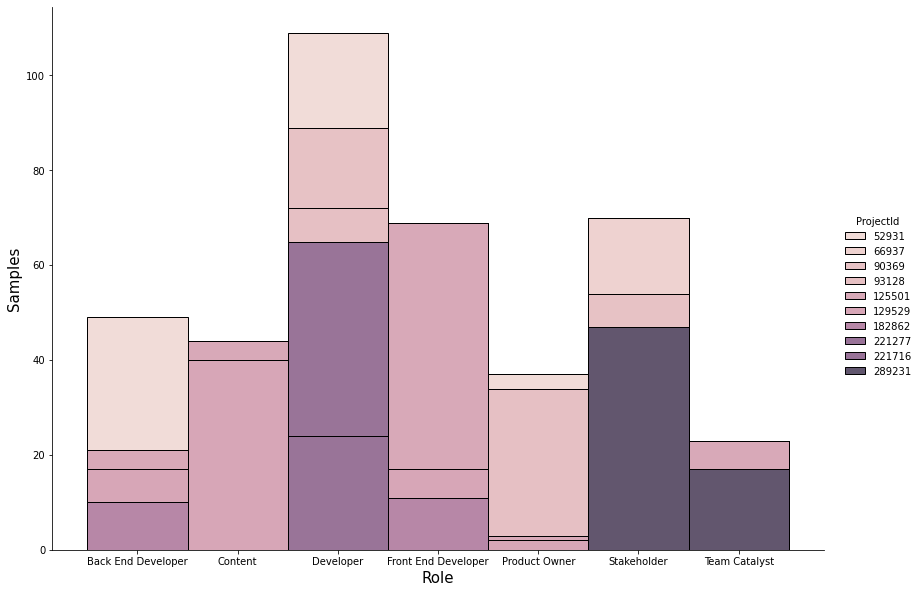}
  \caption{Roles distribution (validation set)}
  \label{RD}
\end{figure*}

\textbf{Front-end Developer:} This role refers to the members who focus on the UI/UX designing and client side applications.

\textbf{Back-end Developer:} This role refers to the members who specifically work on the code/database implementation and server side scripting.

\textbf{Developer:} The developer is categorized as a full stack developer who is proficient in both front-end and back-end development.

\textbf{Product Owner:} A product owner refers to a person who checks the feasibility and key caveats of new user stories, tasks, and issues and adds them to the product backlog.

\textbf{Team Catalyst:} Team catalyst refers to the project managers or similar roles who regulate the project meetings and facilitate team members to ensure that a consistent project velocity is maintained.

\textbf{Content:} The content role refers to the writers and content creators.

\textbf{Stakeholder:} Although the role stakeholder may include all of the above roles, we categorize this role as the external member who is not a part of the development team.

\subsection{Implementation}
The TaskAllocator is implemented in the Python language as it provides actively maintained and well-formed machine learning libraries such as tensorflow\footnote{\url{https://www.tensorflow.org/}}, scikit-learn\footnote{\url{https://scikit-learn.org/stable/}}, and keras. We initially divided the dataset into a train set and a held-out validation set (67/33 split) in order to evaluate the performance of our model.

\subsubsection{Training process}
The pretrained vectors mentioned in Section~\ref{PT} were used as a separate embedding layer while structuring the model's architecture.
The tokenized sequences obtained from the tokenizer were treated as features for our used model while roles as their corresponding labels in the training process. The models were trained on the train set of features and labels. 

\subsubsection{Validation process}
In the validation process, we evaluated the performance of our LSTM model on the validation set. We used K-fold cross validation in order to select the best model. In K-fold cross validation, the training set was split into K folds. The model is trained on (K-1) folds and tested on the remaining fold and repeated K times in order to evaluate the model's performance on unseen data. In training, the model improves to learn by reducing the training loss after each epoch\footnote{An epoch refers to the iteration when each sample in the train set has participated in the learning of the model.}. Early stopping of the model training is also implemented to stop the model from further training if the training accuracy has not improved after few epochs. Once the best model is selected through cross validation, it is trained on the the entire training set and evaluated on a held-out validation set in order to see the LSTM model's performance on individual projects.

\section{Benchmark study}\label{BS}
We further conducted a benchmark study in order to compare the performance of our LSTM model with other alternative models well-known for the text classification problem. As alternative, we have considered two NN models: an ensemble of universal sentence encoder (USE)~\cite{Cer2018} and NN, and convolution neural network 1D (CNN)~\cite{Jacovi2019}. For conventional ML models, we employed: 1) multinomial naive bayes (MNB), 2) support vector classification (SVC), 3) cosine similarity (CS), 4) logistic regression (LR), and 5) random forest (RF). Note that we chose these variants instead of others as they have shown substantial performance in addressing multi-class classification problems in literature~\cite{Anvik2011,Xuan2015,Jonsson2016}. Although the training process for NN models and conventional ML models used in this study vary slightly, we ensured to use similar hyper-parameters and configurations for the training purposes. 
%This benchmark study would help us understand the similarities and differences between the employed LSTM model and other NN and ML models for text classification. 

\subsection{Neural network models}

\begin{figure}[htbp]
 \centering
  \includegraphics[width=0.75\linewidth]{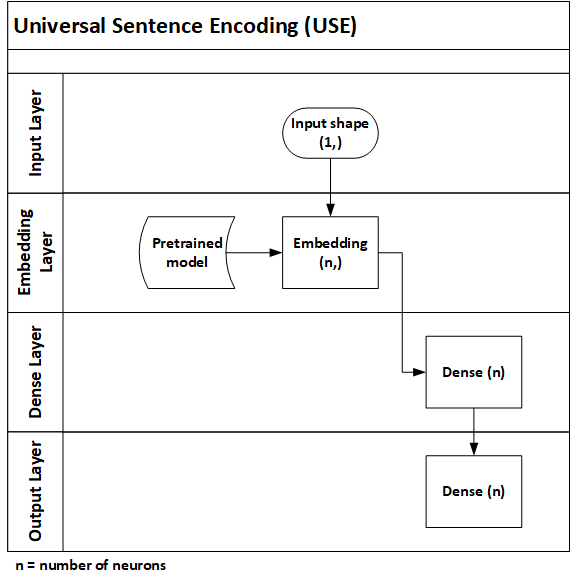}
  \caption{Architecture (USE)}
  \label{USE}
\end{figure}

The first variant of NN we employ in our benchmark study is an ensemble of NN and USE. USE -- developed by Google -- is a text classification model, which is able to efficiently capture sequence of words in a sentence and store its semantic meaning. Fig.~\ref{USE} shows the USE architecture implemented for the benchmark study. The architecture comprises of four layers. 1) The first layer comprises of the input encoded text, which is fed to the embedding layer. 2) The embedding layer is made up of a pre-trained universal sentence encoder and the input of the previous layer. 3) The embedding layer is then followed by a dense layer of interconnected neurons. 4) Lastly, the output dense layer yields classes (roles).

\begin{figure}[htbp]
 \centering
  \includegraphics[width=1.0\linewidth]{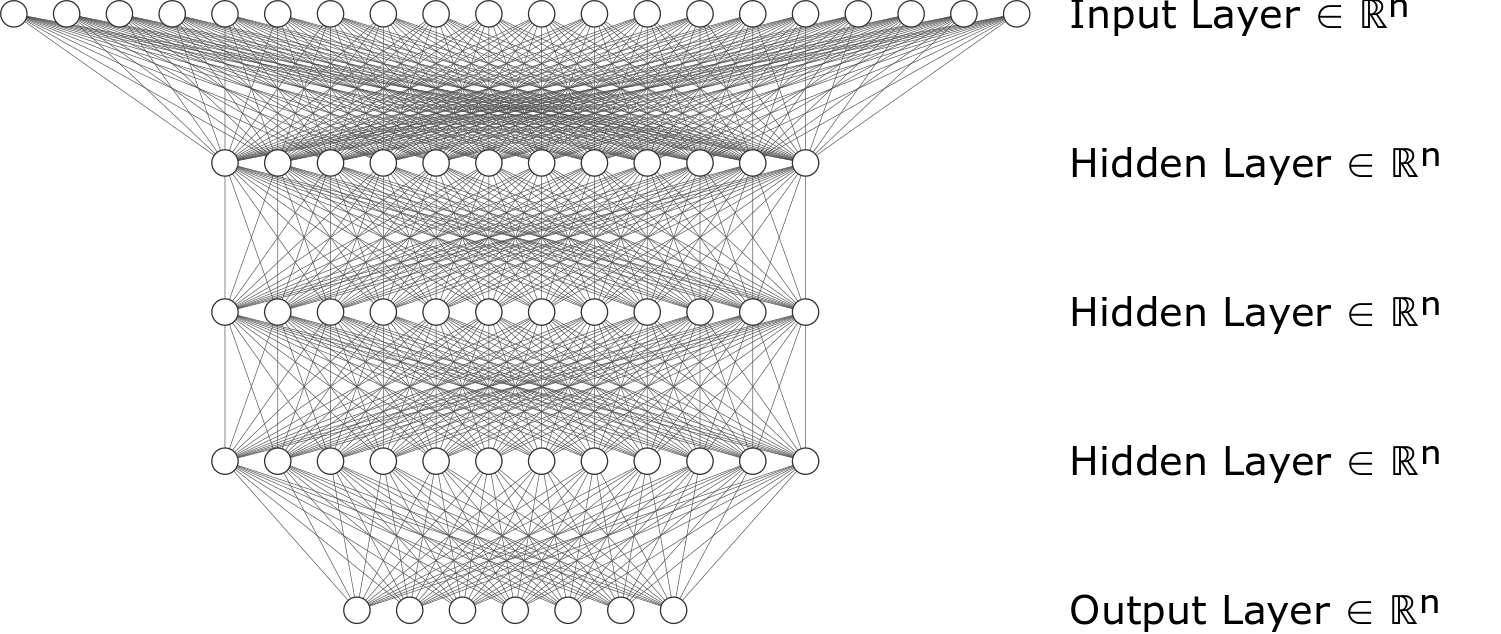}
  \caption{Architecture (CNN)}
  \label{CNN}
\end{figure}

The second NN we employ in our benchmark study is 1-dimensional CNN. We use it instead of multi dimensional CNN due to its promising results shown in text classification~\cite{Jacovi2019}. Especially, the proximity of words can be efficiently captured using 1-dimensional CNN. Fig.~\ref{CNN} shows the layered architecture of the 1D CNN employed in this study representing $n$ number of neurons at each layer.
% The architecture comprises of 1 embedding and input layer (300 neurons), 1D CNN layer (128 neurons each), and an output layer (7 neurons) yielding 7 classes.

\subsection{Conventional machine learning models}

The existing conventional ML models we employed for the benchmark study are: MNB, linear SVC -- an implementation of support vector machine (SVM), CS, LR, and RF. These models have been successfully employed in bug triaging studies in the past~\cite{Anvik2011,Xuan2015,Jonsson2016}. In fact, these ML models rely on the bag of words (BOW) representation. Therefore, we have used term frequency based BOW to represent title as features, which are later used in the training process. 

The first ML model we considered in our benchmark study is called multinomial naive bayes (MNB). MNB computes the class probabilities using Bayes rules~\cite{Kibriya2004}. Kibriya et al.~\cite{Kibriya2004} evaluated four versions of MNB for text categorization and analyzed the performance of each model. Apparently, MNB with term frequency-inverse document frequency (TF-IDF) outperformed the rest when evaluated on the datasets.

Support vector machines are usually applied to binary classification problems, however, its variant known as linear-support vector classification (SVC) has shown better performance in comparison when applied in the textual context~\cite{Xu2009}. The principle behind SVC is the identification of the hyper-plane in the {\it n}-dimensional space, which can distinguish between data points significantly better allowing it to deal with multi class classification problems.

The third ML model we considered in our benchmark study is called cosine similarity (CS) and is used to identify the semantic similarity on words after converting them in TF-IDF weights. These weights describe the word specificity beginning from the first segment of the text~\cite{Mihalcea2006}.

The fourth ML model we considered in our benchmark study is called logistic regression (LR). LR is a linear classifier used to measure the relationship between dependent categorical variable and one or more independent variables. It generates probability estimates using the logistic function. 
%The mathematical representation of the function for logistic regression is shown by Schmidt et al.~\cite{Schmidt2017} .

The last ML model we considered in our benchmark study is called random forest (RF). It is a well-known ML algorithm based on combination of classification trees. An improved version of RF has been proposed by Xu et al.~\cite{Xu2012a}, which is tailored specifically for text categorization.

\subsection{Results and discussion}

We now compare the performance of our employed LSTM model with two other NN models and five conventional ML models, all well-known for the text classification problem. Table~\ref{Evaluationmetrics} summarizes the results of the benchmark study. Overall, results of the study show that the TaskAllocator employing the LSTM model without pre-trained data performed relatively better than other alternatives. The loss metric represents how wrong were the predictions made by the model (summation of errors made by the model for each observation of the validation set) at each epoch during the training process. The lower the loss value, the better the model's capability to perform over the unseen data. Note that the loss metric is only shown for the NN based models due to their iterative learning nature. The accuracy metric represents the percentage of correct predictions made by the model. As shown in Table~\ref{Evaluationmetrics}, the LSTM model showed a prediction accuracy of 69.3\% followed by the USE model with 54.4\% of prediction accuracy whereas 1D CNN has performed the lowest with a prediction accuracy of 42.4\%. The comparable decline in 1D CNN is due to the fact that it was focusing on just the proximity of words and was unable to well identify the high level semantic meaning behind the words in the sentence. Note that USE comes with a universally pre-trained vector by default, therefore, the cell for ``Not Pre-trained'' for USE is set to N/A. Similarly, no pre-training and loss metric was employed in case of conventional ML models, thus N/A. 

\begin{figure}[htbp]
  \centering
  \includegraphics[width=1.0\linewidth]{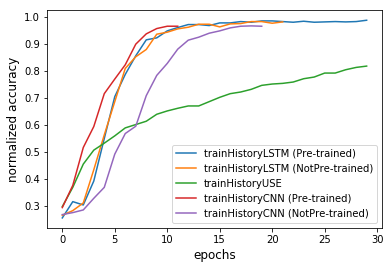}
  \caption{NN models comparison (train set accuracy over 30 epochs)}
  \label{ModelComparison}
\end{figure}

Fig.~\ref{ModelComparison} shows the accuracy of LSTM and other NN during the training process over 30 epochs. The figure shows how well the models were able to learn from the training data over the course of 30 epochs. All the model variants eventually reached a similar training accuracy except USE+NN, which ultimately results in the decrease in the performance on the validation set.

\begin{figure}
  \centering
  \includegraphics[width=1.0\linewidth]{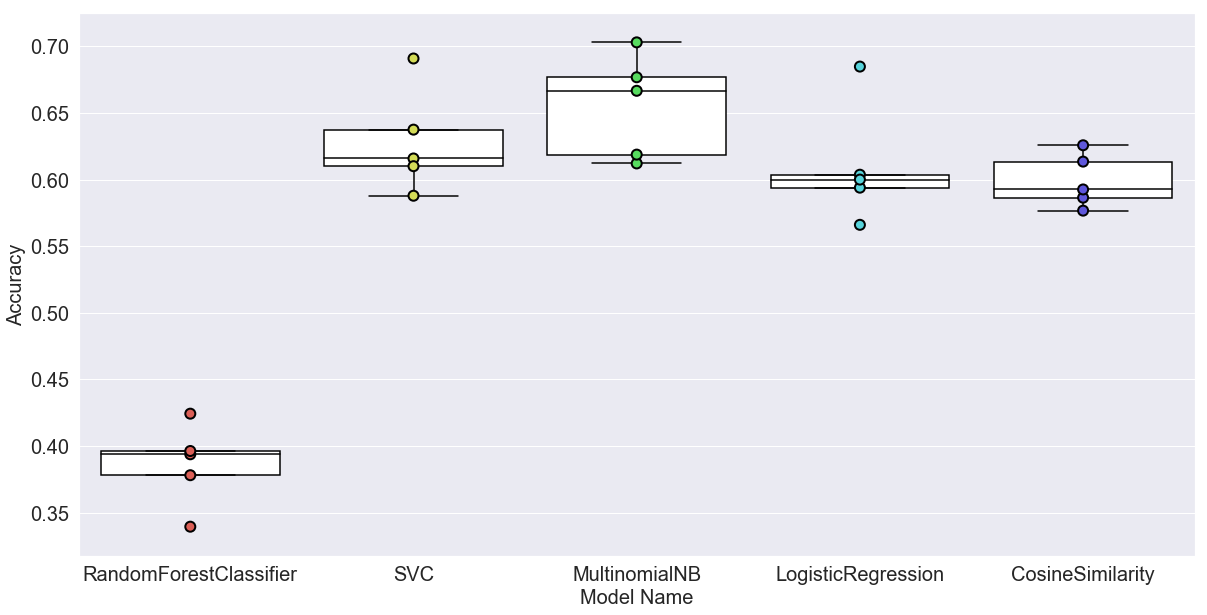}
  \caption{Conventional ML models comparison (train set)}
  \label{TradMLComp}
\end{figure}

On the other hand, the conventional ML models performed on par with the two NN variants with slightly less accuracy. Fig.~\ref{TradMLComp} shows the accuracy of the conventional ML models obtained from K-fold cross validation.
In conventional ML models, MNB outperformed rest of the ML models with an accuracy of 66.3\% followed by SVC and LR with an accuracy of 65\% and 61.8\%, respectively.
 
% Among the conventional ML models, MNB model outperformed the rest with the accuracy of 66.3\%.

We also analysed the models' performance with respect to individual projects. As shown in Table~\ref{Evaluationmetricsbyprojects}, the LSTM model without pre-trained data performed well for all projects except the project ``221716'' (50\%), which is due to the fact that the model was unable to predict half of the actual roles present in the validation set. This could indicate the indistinguishable features on which the model was trained and less number of train samples. A similar behavior has been shown by the pre-trained LSTM model for the same project, which implies that the vocabulary of the pre-trained data was not helpful in case of this project. On the other hand, CNN model showed converse results as the model performed well when pre-trained data is employed except for two projects ``90369'' and ``221716''. The reason for this could be the underlying structure of LSTM and CNN models in capturing features in the training process. The USE+NN model, however, performed relatively moderate in all of the projects when compared to the CNN model and the LSTM model.

Table~\ref{Evaluationmetricsbyprojects} illustrates the performance (in terms of accuracy) of the other two variants of NN and the conventional ML models with respect to projects. Table~\ref{Evaluationmetricsbyprojects} further shows the results of the models with and without pre-trained data. The detail regarding how the pre-trained data employed in the training process is explained in Section~\ref{PT}. One of the interesting observations we made while conducting the benchmark study was the volatile behavior of the RF model. While for few projects, such as ``221277'' and ``221716'', RF achieved an accuracy of 100\% whereas for project ``182862'' the RF model was unable to correctly predict any sample in the validation set. This is presumably due to the lower number of samples in the train set, which made it difficult for the RF model to find highly discriminatory textual features from the sub-samples.

To summarize the results, the recommendations made by the TaskAllocator with an accuracy of $\sim$70\% infer that, at least, 2 out of 3 predictions made by the TaskAllocator are correct, thus helping project managers make intelligible tasks allocations. As the TaskAllocator is independent of the underlying model, its performance could be further enhanced by employing newly developed ML models in the future.

% WHAT I"M MISSING HERE IS: WHAT ARE THE PRACTICAL IMPLICATIONS FOR THE PROJECT MANAGER: how good are the results, how often would he/she get a wrong recommendations. What do the numbers of loss and accuracy mean in practical terms for the manager?!

\begin{table}
\centering
\caption{Evaluation metrics (Loss \& Accuracy) [Validation set]}
\begin{tabular}{|c|c|c|c|c|}
\hline
Model & \multicolumn{2}{c|}{Loss}& \multicolumn{2}{c|}{Accuracy} \\
\hline
& P & !P  & P & !P  \\
\hline
LSTM & 1.327 &\textbf{1.240}& \textbf{68.6} & \textbf{69.3}\\
\hline
USE+NN& \textbf{1.309} &N/A& 54.4 & N/A \\
\hline
CNN& 2.669 &2.3928& 37.2 & 42.4 \\
\hline
\hline
MNB& N/A &N/A& N/A & 66.3 \\
\hline
Linear SVC& N/A &N/A& N/A & 65.0\\
\hline
LR& N/A &N/A& N/A & 61.8 \\
\hline
CS& N/A &N/A& N/A & 59.1 \\
\hline
RF& N/A &N/A& N/A & 43.6 \\
\hline
\multicolumn{5}{l}{%
  \begin{minipage}{6.5cm}~\\
    \tiny P = Pre-trained, !P = Not Pre-trained
  \end{minipage}%
}\\
\end{tabular}
\label{Evaluationmetrics}
\end{table}

\begin{table*}
\centering
\caption{Evaluation metric (Accuracy) by projects [Validation set]}
\begin{tabular}{|c|c|c|c|c|c|c|c|c|c|c|}
\hline
Project Id &\multicolumn{2}{c|}{LSTM} & \multicolumn{2}{c|}{CNN} & USE+NN& MNB & SVC & CS & LR & RF \\
% \hline
% &\multicolumn{2}{c|}{Accuracy}&\multicolumn{2}{c|}{Accuracy}&Accuracy&Accuracy&Accuracy&Accuracy&Accuracy&Accuracy\\
\hline
&P&!P&P&!P&P&&&&&\\
\hline
221277&82.9 & 92.6 & 78 &70.7 &78 &90.2 &90.2 & 75.6 &97.5 & \textbf{100} \\
\hline
289231& \textbf{67.1} & 64 & 39  &43.7 & 43.7 & 62.5&56.2 &62.5 & 46.8 & 23.4 \\
\hline
221716&45.8 & 50 & 41.6  &29.1 &37.5 &54.1 &58.3 &54.1 &66.6 & \textbf{100} \\
\hline
66937 &68.7 &\textbf{87.5}  & 25 &56.2 &62.5 &81.2 &87.5 &81.2 & \textbf{87.5} &31.2 \\
\hline
129529& 50.9 &\textbf{61.8} & 52.7 &54.5 &50.9 & \textbf{61.8} &50.9 &60.0 &36.3 & 9.09 \\
\hline
93128 &65.7 & 68.4 & 52.6 & 65.7  &57.8  & 68.4& \textbf{73.6}& 47.3&71.0 & 34.2 \\
\hline
90369 &62.5 &62.5 & 41.6 & 41.6  &50 &45.8 &54.1 &37.5 &54.1 &\textbf{70.8} \\
\hline
52931 &60.7 &\textbf{72.5}  & 39.2 & 50.9  &45 & 66.6& 66.6&74.5 &68.6 &50.9 \\
\hline
125501&65.6 &\textbf{68.6} & 56.7 & 46.2  &62.6 &67.1 & 64.1&47.7 & 61.1 & 43.2 \\
\hline
182862&47.6 &61.9 & 38 &57.1 &57.1 &61.9 &\textbf{66.6} &47.6 &57.1 & 0.0 \\
\hline
\multicolumn{11}{l}{%
  \begin{minipage}{6.5cm}~\\
    \tiny P = Pre-trained, !P = Not Pre-trained\\
    \tiny top models with accuracy for each project are emphasized
  \end{minipage}%
}\\
\end{tabular}
% \begin{tablenotes}
% \item[1] P = Pre-trained, !P = Not Pre-trained
% \end{tablenotes}
\label{Evaluationmetricsbyprojects}
\end{table*}

%% file: usability.tex
\subsection{TaskAllocator prototype}
\begin{figure}[htbp]
  \centering
  \includegraphics[width=0.75\linewidth]{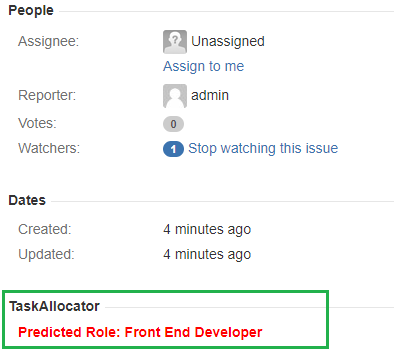}
  \caption{The TaskAllocator plugin}
  \label{PluginSS}
\end{figure}

In order to demonstrate the applicability of the TaskAllocator, we developed an install-able JIRA plugin. The plugin recommends a potentially suitable role whenever a reporter (product owner) adds a new task in the project. The plugin output is shown in the view issue page. A screen shot of the plugin is shown in Fig.~\ref{PluginSS}. The source code of the plugin and steps to integrate it with JIRA are available on Github.\footnote{\url{https://github.com/jku-isse/TaskAllocator}}

\subsection{Dataset}
In order to further the research, the extracted dataset has also been made publicly available.\footnote{\url{https://github.com/jku-isse/TaskAllocator/tree/master/Dataset}} The dataset contains \textit{.csv} files for individual projects and comprises of fields with headers named as \textit{ProjectId}, \textit{ProjectName}, \textit{Title} of the task, \textit{Description} of the task, and the \textit{Role} of the assignee. Due to the scarcity of the publicly available data in this regard, this dataset could help researchers to further conduct studies on understanding the semantic meaning of tasks by analysing the titles and descriptions, and their association with the designated roles. It can also greatly benefit the scientific community for performing social network analysis by extracting emotional aspects from textual features and identifying a co-relation with roles.

\subsection{Further implications}
Although the shown plugin has been developed for the JIRA platform, the TaskAllocator approach is generic and platform independent. The TaskAllocator helps project managers using any project management or issue tracking platform to be aware of their project history by providing them the capability to identify most suitable and befitting roles in their distributed teams for an incoming task, which is crucial for sprint planning. For researchers, this study helps understanding whether the textual features of tasks can be an effective predictor in identifying coarse-grained roles in a team.
%For managers, this study would also help them to better orient their teams and facilitate them in correctly identifying potential roles for the future tasks earlier in their sprint planning.

%% file: ThreatsToValidity.tex
\subsection{External validity}
This study is prone to the external validity threat due to the fact that the TaskAllocator can not be generalized to all projects as the vocabulary and distribution of roles may vary. To overcome this threat, we performed K-fold cross validation in order to select the best model. Also, when using a trained model for other projects, we introduced a preference mechanism on top of the model, which selects the most suitable role from a given set of roles present in a project (only valid in cross-project setting).

\subsection{Internal validity}
The internal validity threat arises due to the execution of a weak research protocol. We reduced this threat by following the defined project criteria when extracting data from the repository. In order to remove the model's bias, we followed the standard model validation practice established in literature. We also evaluated the model on individual projects' validation set and used a well-established evaluation metric to determine its performance.

\subsection{Construct validity}

The overall goal of this study is to determine which role is suitable for an incoming task. As aforementioned, most project management platforms do not provide the roles for the team members in their public projects. To overcome this, we chose {\it Taiga.io} as our primary repository for obtaining public projects, which has quite extensive list of projects with explicitly mentioned roles of the team. However, the lack of relatively less data implies that the proposed results still need to be evaluated on a comparatively larger dataset. 

\subsection{Conclusion validity}
The conclusion validity refers to the authenticity and correctness of the obtained results. In order to overcome this threat, we followed industrially recognized standard practice by employing K-fold cross validation, which not only removes model's bias towards the specific data but also helps in drawing an accurate conclusion. We further introduced a preference mechanism, which eases the role selection and adds the flexibility of selecting the best role from within a project when using a model trained on a different project's data.

%% file: conclusion.tex
The contribution of this paper is two fold. First,  we propose a recommendation approach -- called the TaskAllocator -- to solve the task allocation problem at a coarse-grained level. The TaskAllocator leverages the LSTM model in order to recommend the most befitting role for an incoming task. This scheme is useful when developers are either over-committed or have already changed the team/role, or when the manager wants to distribute the task knowledge among multiple team members. We develop a corresponding JIRA plugin in order to demonstrate the prediction capability and applicability of the TaskAllocator. The source code of the plugin and the employed dataset has been made publicly available for the researchers. One of the distinctive features of the TaskAllocator is its flexible architecture. Currently, our recommendation approach is based on the LSTM model due to its superior performance over its contemporaries. However, the TaskAllocator is flexible enough to accommodate any ML/NN model that demonstrates an improved performance in the future.

Second, we perform a benchmark study in order to determine the performance of the TaskAllocator by comparing it with contemporary NN and ML models. The results of the benchmark study show that the employed LSTM model outperformed its alternatives with an overall accuracy of 69.3\%, thanks to its ability to remember long sequences. The multinomial naive bayes model -- a well-known ML model for addressing multi-class text classification problems -- showed an accuracy of 66.3\%, which is slightly less than than the performance of the LSTM model.

In the future, we intend to enlarge the dataset by accompanying other projects for a better model training process. We also intend to evaluate other text classification models such as FastText\footnote{\url{https://fasttext.cc/}} and StarSpace\footnote{\url{https://github.com/facebookresearch/StarSpace}}. Apart from using online project management tools and platforms, we also aim to use the TaskAllocator in industrial problems.

%% file: main.bbl
\begin{thebibliography}{10}
\providecommand{\url}[1]{#1}
\csname url@samestyle\endcsname
\providecommand{\newblock}{\relax}
\providecommand{\bibinfo}[2]{#2}
\providecommand{\BIBentrySTDinterwordspacing}{\spaceskip=0pt\relax}
\providecommand{\BIBentryALTinterwordstretchfactor}{4}
\providecommand{\BIBentryALTinterwordspacing}{\spaceskip=\fontdimen2\font plus
\BIBentryALTinterwordstretchfactor\fontdimen3\font minus
  \fontdimen4\font\relax}
\providecommand{\BIBforeignlanguage}[2]{{%
\expandafter\ifx\csname l@#1\endcsname\relax
\typeout{** WARNING: IEEEtran.bst: No hyphenation pattern has been}%
\typeout{** loaded for the language `#1'. Using the pattern for}%
\typeout{** the default language instead.}%
\else
\language=\csname l@#1\endcsname
\fi
#2}}
\providecommand{\BIBdecl}{\relax}
\BIBdecl

\bibitem{Alzoubi2015d}
\BIBentryALTinterwordspacing
Y.~I. Alzoubi, A.~Q. Gill, and A.~Al-Ani, ``{Empirical studies of
  geographically distributed agile development communication challenges: A
  systematic review},'' \emph{Information {\&} Management}, vol.~53, no.~1,
  pp.~--, 2015. [Online]. Available:
  \url{http://www.sciencedirect.com/science/article/pii/S0378720615000877}
\BIBentrySTDinterwordspacing

\bibitem{Shrivastava2015}
\BIBentryALTinterwordspacing
S.~V. Shrivastava and U.~Rathod, ``{Categorization of risk factors for
  distributed agile projects},'' \emph{Information and Software Technology},
  vol.~58, pp. 373--387, 2015. [Online]. Available:
  \url{http://dx.doi.org/10.1016/j.infsof.2014.07.007}
\BIBentrySTDinterwordspacing

\bibitem{Aslam2018}
W.~Aslam and F.~Ijaz, ``{A Quantitative Framework for Task Allocation in
  Distributed Agile Software Development},'' \emph{IEEE Access}, vol.~6, pp.
  15\,380--15\,390, 2018.

\bibitem{Imtiaz2017}
S.~Imtiaz and N.~Ikram, ``{Dynamics of task allocation in global software
  development},'' \emph{Journal of Software: Evolution and Process}, vol.~29,
  no.~1, pp. 1--17, 2017.

\bibitem{Barcus2008}
A.~Barcus and G.~Montibeller, ``{Supporting the allocation of software
  development work in distributed teams with multi-criteria decision
  analysis},'' \emph{Omega}, vol.~36, no.~3, pp. 464--475, 2008.

\bibitem{Falcini2017}
F.~Falcini, G.~Lami, I.~Science, A.~M. Costanza, and F.~C. Automobiles, ``{Deep
  Learning in Automotive Software},'' \emph{IEEE Software}, vol.~34, no.~3, pp.
  56--63, 2017.

\bibitem{Chen2018b}
\BIBentryALTinterwordspacing
C.~Chen, O.~Li, C.~Tao, A.~J. Barnett, J.~Su, and C.~Rudin, ``{This Looks Like
  That: Deep Learning for Interpretable Image Recognition},'' in \emph{33rd
  Conference on Neural Information Processing Systems (NeurIPS 2019)}, no.
  NeurIPS, 2018, pp. 1--12. [Online]. Available:
  \url{http://arxiv.org/abs/1806.10574}
\BIBentrySTDinterwordspacing

\bibitem{Choetkiertikul2019}
M.~Choetkiertikul, H.~K. Dam, T.~Tran, T.~Pham, A.~Ghose, and T.~Menzies, ``{A
  Deep Learning Model for Estimating Story Points},'' \emph{IEEE Transactions
  on Software Engineering}, vol.~45, no.~7, pp. 637--656, 2019.

\bibitem{Shafiq2020a}
S.~Shafiq, C.~Mayr-Dorn, A.~Mashkoor, and A.~Egyed, ``{Towards Optimal Assembly
  Line Order Sequencing with Reinforcement Learning: A Case Study},''
  \emph{IEEE International Conference on Emerging Technologies and Factory
  Automation, ETFA}, vol. 2020-Septe, pp. 982--989, 2020.

\bibitem{Anvik2006}
J.~Anvik, L.~Hiew, and G.~C. Murphy, ``{Who should fix this bug?}'' in
  \emph{Proceedings - International Conference on Software Engineering}, 2006,
  pp. 361--370.

\bibitem{Mani2018}
S.~Mani, A.~Sankaran, and R.~Aralikatte, ``{DeepTriage : Exploring the
  Effectiveness of Deep Learning for Bug Triaging},'' in \emph{Proceedings of
  the ACM India Joint International Conference on Data Science and Management
  of Data}, 2019, pp. 171--179.

\bibitem{Systems2018}
S.~C. J and A.~Mahendran, ``{Automated Bug Assignment in Software Maintenance
  Using Graph Databases},'' \emph{I.J. Intelligent Systems and Applications},
  vol.~2, no. February, pp. 27--36, 2018.

\bibitem{Choquette-choo2019}
C.~A. Choquette-choo, D.~Sheldon, J.~Proppe, J.~Alphonso-gibbs, and H.~Gupta,
  ``{A multi-label , dual-output deep neural network for automated bug
  triaging},'' in \emph{18th IEEE International Conference On Machine Learning
  And Applications (ICMLA)}, 2019, pp. 1--8.

\bibitem{Yadav2019}
\BIBentryALTinterwordspacing
A.~Yadav, S.~Kumar, and J.~S. Suri, ``{Ranking of software developers based on
  expertise score for bug triaging},'' \emph{Information and Software
  Technology}, vol. 112, no. February, pp. 1--17, 2019. [Online]. Available:
  \url{https://doi.org/10.1016/j.infsof.2019.03.014}
\BIBentrySTDinterwordspacing

\bibitem{DeMelo2013}
\BIBentryALTinterwordspacing
C.~O. {De Melo}, D.~{S. Cruzes}, F.~Kon, and R.~Conradi, ``{Interpretative case
  studies on agile team productivity and management},'' \emph{Information and
  Software Technology}, vol.~55, no.~2, pp. 412--427, 2013. [Online].
  Available: \url{http://dx.doi.org/10.1016/j.infsof.2012.09.004}
\BIBentrySTDinterwordspacing

\bibitem{Baysal2009}
O.~Baysal, M.~W. Godfrey, and R.~Cohen, ``{A bug you like: A framework for
  automated assignment of bugs},'' \emph{IEEE International Conference on
  Program Comprehension}, pp. 297--298, 2009.

\bibitem{Asri2018a}
I.~E. Asri, M.~Abdou, and J.~Idrissi, ``{Understanding OSS Project's
  Collaborative Dynamics: Core and Peripheral Interactions},''
  \emph{International Journal of Scientific {\&} Engineering Research}, vol.~9,
  no.~11, pp. 1541--1551, 2018.

\bibitem{Moe2013}
N.~B. Moe, ``{Key challenges of improving agile teamwork},'' \emph{Lecture
  Notes in Business Information Processing}, vol. 149, no. June 2013, pp.
  76--90, 2013.

\bibitem{belling2020agile}
S.~Belling, ``{Agile Teams and Challenges},'' in \emph{Succeeding with Agile
  Hybrids}.\hskip 1em plus 0.5em minus 0.4em\relax Springer, 2020, pp. 63--72.

\bibitem{Anvik2011}
J.~Anvik and G.~C. Murphy, ``{Reducing the effort of bug report triage:
  Recommenders for development-oriented decisions},'' \emph{ACM Transactions on
  Software Engineering and Methodology}, vol.~20, no.~3, pp. 1--35, 2011.

\bibitem{Pichler2007}
H.~Pichler, ``{Be successful , take a hostage or “ outsourcing the
  outsourcing Manager ”},'' in \emph{International Conference on Global
  Software Engineering}, 2007, pp. 156--161.

\bibitem{Cetin2020}
H.~A. {\c{C}}etin, ``{Identifying Key Developers using Artifact Traceability
  Graphs},'' in \emph{16th ACM International Conference on Predictive Models
  and Data Analytics in Software Engineering}, 2020, pp. 1--10.

\bibitem{Ahmad2011}
A.~Ahmad, \emph{{Effective Distribution of Roles and Responsibilities in Global
  Software Development Teams (Master Thesis)}}.\hskip 1em plus 0.5em minus
  0.4em\relax Blekinge Institute of Technology, 2012.

\bibitem{Lings2007}
B.~Lings, B.~Lundell, P.~J. {\AA}gerfalk, and B.~Fitzgerald, ``{A reference
  model for successful Distributed Development of Software Systems},'' in
  \emph{International Conference on Global Software Engineering}, 2007, pp.
  130--139.

\bibitem{Efstathiou2018}
V.~Efstathiou, C.~Chatzilenas, and D.~Spinellis, ``{Word embeddings for the
  software engineering domain},'' \emph{Proceedings - International Conference
  on Software Engineering}, pp. 38--41, 2018.

\bibitem{Silva-Palacios2017}
D.~Silva-Palacios, C.~Ferri, and M.~J. Ram{\'{i}}rez-Quintana, ``{Improving
  Performance of Multiclass Classification by Inducing Class Hierarchies},''
  \emph{Procedia Computer Science}, vol. 108, pp. 1692--1701, 2017.

\bibitem{Shafiq2017}
S.~Shafiq and I.~Inayat, ``{Towards studying the communication patterns of
  Kanban teams: A research design},'' in \emph{Proceedings - 2017 IEEE 25th
  International Requirements Engineering Conference Workshops, REW 2017}, 2017,
  pp. 303--306.

\bibitem{Shafiq2019}
------, ``{Communication Patterns of Kanban Teams and their Impact on Iteration
  Performance and Quality},'' in \emph{2019 45th Euromicro Conference on
  Software Engineering and Advanced Applications (SEAA)}.\hskip 1em plus 0.5em
  minus 0.4em\relax IEEE, 2019, pp. 164--168.

\bibitem{Lantz2019}
B.~Lantz, \emph{{Machine learning with R: expert techniques for predictive
  modeling}}.\hskip 1em plus 0.5em minus 0.4em\relax Packt Publishing Ltd,
  2019.

\bibitem{Cer2018}
D.~Cer, Y.~Yang, S.~yi~Kong, N.~Hua, N.~Limtiaco, R.~{St. John}, N.~Constant,
  M.~Guajardo-C{\'{e}}spedes, S.~Yuan, C.~Tar, Y.~H. Sung, B.~Strope, and
  R.~Kurzweil, ``{Universal sentence encoder for English},'' \emph{EMNLP 2018 -
  Conference on Empirical Methods in Natural Language Processing: System
  Demonstrations, Proceedings}, pp. 169--174, 2018.

\bibitem{Jacovi2019}
A.~Jacovi, O.~{Sar Shalom}, and Y.~Goldberg, ``{Understanding Convolutional
  Neural Networks for Text Classification},'' in \emph{Proceedings of the 2018
  EMNLP Workshop BlackboxNLP: Analyzing and Interpreting Neural Networks for
  NLP}.\hskip 1em plus 0.5em minus 0.4em\relax Brussels, Belgium: Association
  for Computational Linguistics, 2018, pp. 56--65.

\bibitem{Xuan2015}
J.~Xuan, H.~Jiang, Y.~Hu, Z.~Ren, W.~Zou, Z.~Luo, and X.~Wu, ``{Towards
  effective bug triage with software data reduction techniques},'' \emph{IEEE
  Transactions on Knowledge and Data Engineering}, vol.~27, no.~1, pp.
  264--280, 2015.

\bibitem{Jonsson2016}
\BIBentryALTinterwordspacing
L.~Jonsson, M.~Borg, D.~Broman, K.~Sandahl, S.~Eldh, and P.~Runeson,
  \emph{{Automated bug assignment: Ensemble-based machine learning in large
  scale industrial contexts}}.\hskip 1em plus 0.5em minus 0.4em\relax Empirical
  Software Engineering, 2016, vol.~21, no.~4. [Online]. Available:
  \url{http://dx.doi.org/10.1007/s10664-015-9401-9}
\BIBentrySTDinterwordspacing

\bibitem{Kibriya2004}
A.~M. Kibriya, E.~Frank, B.~Pfahringer, and G.~Holmes, ``{Multinomial naive
  bayes for text categorization revisited},'' \emph{Lecture Notes in Artificial
  Intelligence (Subseries of Lecture Notes in Computer Science)}, vol. 3339,
  pp. 488--499, 2004.

\bibitem{Xu2009}
Y.~Xu, Y.~Shao, Y.~Tian, and N.~Deng, ``{Linear Multi-class Classification
  Support Vector Machine},'' in \emph{International Conference on Multiple
  Criteria Decision Making}.\hskip 1em plus 0.5em minus 0.4em\relax Berlin,
  Heidelberg: Springer, 2009, pp. 635--642.

\bibitem{Mihalcea2006}
R.~Mihalcea, C.~Corley, and C.~Strapparava, ``{Corpus-based and knowledge-based
  measures of text semantic similarity},'' \emph{Proceedings of the National
  Conference on Artificial Intelligence}, vol.~1, pp. 775--780, 2006.

\bibitem{Xu2012a}
B.~Xu, X.~Guo, Y.~Ye, and J.~Cheng, ``{An improved random forest classifier for
  text categorization},'' \emph{Journal of Computers (Finland)}, vol.~7,
  no.~12, pp. 2913--2920, 2012.

\end{thebibliography}
